\documentclass[a4paper]{jpconf}
\usepackage{graphicx}
\usepackage{amsmath}
\begin{document}
\title{Astrophysical magnetohydrodynamical outflows in the extragalactic binary system LMC X-1}

\author{Th V Papavasileiou$^{1,2,a}$, D A Papadopoulos$^{1,b}$ 
and T S Kosmas$^{1,c}$}

\address{$^1$Division of Theoretical Physics, University of Ioannina, GR-45110 Ioannina, Greece}
\address{$^2$Department of Informatics, 
University of Western Macedonia, GR-52100 Kastoria, Greece}
%\address{$^3$ Department of Computer Science, 
%University College London, Gower Street, London WC1E~6BT, UK}

\ead{$^a$theodora836@gmail.com, $^b$mitsospapad@hotmail.com, $^c$hkosmas@uoi.gr}

%\corref{T.K}}

%\cortext[cor]{Corresponding author}

% \address{Division of Theoretical Physics, University of Ioannina,GR-45110 Ioannina, Greece}

%\cortext[T.K]{Corresponding author}

%\author{Jacky Mucklow}

%\address{Production Editor, \jpcs, \iopp, Dirac House, Temple Back, Bristol BS1~6BE, UK}

%\ead{jacky.mucklow@iop.org}
\vskip 1em
\hspace{5 em}
 \vskip 0em \hspace{5 em} LMC X-1, $\gamma$-ray emission  
  
\begin{abstract}
In this work, at first we present a model of studying astrophysical flows of binary systems 
and microquasars based on the laws of relativistic magnetohydrodynamics. Then, by solving 
the time independent transfer equation, we estimate the primary and secondary particle 
distributions within the hadronic astrophysical jets as well as the emissivities of high 
energy neutrinos and  $\gamma$-rays. One of our main goals is, by taking into consideration 
the various energy-losses of particles into the hadronic jets, to determine through the 
transport equation the respective particle distributions focusing on relativistic hadronic 
jets of binary systems. As a concrete example we examine the extragalactic binary system 
LMC X-1 located in the Large Magellanic Cloud, a satellite galaxy of our Milky Way Galaxy.   
\end{abstract}

%\begin{keyword}
%\end{keyword}

\section{Introduction}

In recent years, astrophysical magnetohydrodynamical flows in Galactic and 
extragalactic X-ray binary systems
and microquasars have been modelled with the purpose of studying their 
multi messenger emissions (radiative multiwavelength emission and particle, 
e.g. neutrino, emissions) \cite{Romero,Reynoso}. 
For the detection of such emissions, extremely sensitive detector 
tools are in operation for recording their signals reaching the Earth
like KM3NeT, IceCube, ANTARES, etc.
Modelling offers good support for future attempts to detect them 
while in parallel several numerical simulations have been performed towards 
this aim \cite{Ody-Smpon-2015,Ody-Smpon-2017,Ody-Smpon-2018}.

In general, the astrophysical jets, and specifically those coming from microquasars, 
may well be described as fluid flow emanating from the vicinity of the 
compact object of the binary system. We mention that, microquasars 
are binary systems consisted of a compact stellar object and a donor 
(companion) star. Well known microquasar systems include the Galactic 
X-ray binaries SS433, Cyg X-1, etc.,
while from the extragalactic ones we mention the LMC X-1, LMC X-3 (in 
the neighbouring galaxy of the Large Magellanic Cloud) 
\cite{Papav-Ody-Sinat-AHEP,MSc-Papav},
and the Messier X-7 (in the Messier 33 galaxy).
Their respective relativistic jets emit radiation in various wavelength 
bands and high energy neutrinos. 

Up to now, the SS433 is the only microquasar observed with a definite hadronic content 
in its jets, as verified from observations of their spectra (see Ref. 
\cite{Ody-Smpon-2015,Ody-Smpon-2017,Ody-Smpon-2018} and references therein). 
Radiative transfer calculations may be performed at every 
point in the jet, for a range of frequencies (energies), at every location 
\cite{Smponias_tsk_2011}, providing the relevant emission and absorption coefficients. 
Line-of-sight integration, afterwards, provides synthetic images of $\gamma$-ray emission, 
at the energy-window of interest \cite{Smponias_tsk_2011,Smponias_tsk_2014}. 

The relativistic treatment of jets, considers various energy loss mechanisms that occur 
due to several hadronic processes, particle decays and particle scattering
\cite{Romero,Reynoso}. In the known fluid 
approximation, macroscopically the jet matter behaves as a fluid collimated by 
the magnetic field. At a smaller scale, consideration of the kinematics of the jet
plasma becomes necessary for treating shock acceleration effects.
 
In the model employed in this work, the jets are considered to be rather conic 
along the z-axis (ejection axis) with a radius $r(z) = z \tan{\xi}$, where $\xi$ 
its half-opening angle. The jet radius at its base, $r_0$, is given by 
$r_0 = z_0 \tan{\xi}$, where $z_0$ is the distance of the jet's base to the 
central compact object. According to the jet-accretion speculation, only 10\% of 
the system's Eddington luminosity $(L_k=1.2\times 10^{37}M$ $erg/s$, M in solar 
masses$)$ is transferred to the 
jet for acceleration and collimation through the magnetic field given by the 
equipartition of magnetic and kinetic energy density as $B=\sqrt{8\pi\rho _k(z)}$
(see Ref. \cite{Romero,Reynoso,Smponias_tsk_2011,Smponias_tsk_2014}. 

In the hadronic models assumed in this work, a small portion of the hadrons (mainly protons) 
$q_r \approx 0.1$ are accelerated with rate $t_{acc}^{-1}\simeq\eta ceB/E_p$ due to the 
2nd order Fermi acceleration mechanism to nearly relativistic velocities (with $\eta =0.1$
being the acceleration efficiency). That results in a power-law distribution 
given in the jet's rest frame by $N'(E')=K_0E'^{-2}$ $GeV^{-1}cm^{-3}$,
where $K_0$ is a normalization constant. 

\section{Interaction mechanisms and Energy loss rates in hadronic astrophysical jets}

In the recent literature, three are the main interaction mechanisms for relativistic 
protons. These include interaction with a stellar wind, a radiation field composed of 
internal and external emission sources and, finally, the cold hadronic matter of the 
jet. In this work we focus on the last interaction because it dominates over the 
other two. 

The p-p interactions initialise a reaction chain leading finally to neutrino and 
$\gamma$-ray production, that can reach as far as the Earth where they are being 
detected by undersea water and under-ice detectors such as KM3NeT, ANTARES and 
IceCube. The aforementioned reaction chain begins with the inelastic p-p collisions 
of the relativistic protons on the cold ones inside the jet which result to neutral 
($\pi^0$) and charged ($\pi^\pm$) pion production. Neutral pions decay into $\gamma$-ray 
photons while the charged ones decay into muons and neutrinos. Subsequently, muons also 
decay into neutrinos.
These are the main reactions feeding the neutrino and gamma-ray production channel in our 
models. Furthermore, high energy emission spectra can be more emphatically explained by 
leptonic models where the energy is transferred by leptons (electrons) instead of hadrons, 
whereas hadronic models are more suitable for neutrino production processes.

All particles that take part in the neutrino and gamma-ray production processes lose 
energy while travelling along the acceleration zone which could be due to different 
mechanisms discussed below. At first, the particles can be subjected to adiabatic 
energy losses due to jet expansion along the ejection axis with a rate
\begin{equation}
t_{ad}^{-1}=\frac{2\upsilon _{b}}{3z}
\end{equation} 
where $\upsilon _{b}$ is the jet's bulk velocity. Particles can also lose energy because they 
collide with the jet's cold matter with rate given by
\begin{equation}
t_{ip}^{-1}=\frac{1}{2}n(z)c\sigma _{ip}^{inel}(E) \, ,\qquad 
n(z) = \frac{(1-q_r)L_k}{\Gamma m_pc^2\pi r(z)^2\upsilon _{b}}\, .
\end{equation}   
In the latter expression, $n(z)$ denotes the cold protons density and $\Gamma$ is the Lorentz 
factor corresponding to the jet's velocity. The factor $1/2$ is the p-p collision in-elasticity 
coefficient. Also, the $i$-index represents the different particles that take part in the 
collisions with the cold protons. These, due to small muon mass can be mainly relativistic 
protons and pions. The inelastic cross section for the p-p scattering is given in \cite{Kelner} 
which equals the cross section regarding the $\pi -p$ scattering with a 2/3 factor 
\cite{Ody-Smpon-2015,Ody-Smpon-2017,Ody-Smpon-2018}.  

%\begin{equation}
%\sigma _{pp}^{inel}(E_p)=(0.25L^2+1.88L+34.3)\left[1-
%\left(\frac{E_{th}}{E_p}\right)^4\right]^2\times 10^{-27} cm^2
%\end{equation}      
%where $L=ln(E_p/1000)$ with $E_p$ in GeV and $E_{th}=1.2$ $GeV$ the threshold for the production 
%of a single neutral pion. Respectively, for pions it holds \cite{Gaisser}
%\begin{equation}
%\sigma _{\pi p}^{inel}(E_{\pi})\simeq \frac{2}{3}\sigma _{pp}^{inel}(E_{\pi})
%\end{equation} 
In addition, particles accelerated by the magnetic fields emit synchrotron radiation. 
Thus, gradually they lose part of their energy with a rate
\begin{equation}
t_{sync}^{-1}=\frac{4}{3}\left(\frac{m_e}{m}\right)^3\frac{\sigma_TB^2}{8\pi m_ec}\gamma
\end{equation} 
where $\gamma =E/mc^2$ and $\sigma _T=6.65\times 10^{-25}$ $cm^2$, the known Thomson cross 
section. 

Finally, protons and pions can lose energy interacting through X-ray, UV and synchrotron radiation 
due to photo-pion production, while smaller particles such as muons transfer part of their
energy to low-energy photons due to inverse Compton scattering. However, such contributions 
can be ignored compared to those mentioned above. 

\section{ Calculation of the particle distributions } 

In the steady-state model, the jet's particle distributions obey the transfer equation
as \cite{Ody-Smpon-2015,Ody-Smpon-2017,Ody-Smpon-2018}
\begin{equation}
\frac{\partial N(E,z)b(E,z)}{\partial E}+t^{-1}N(E,z)=Q(E,z) \, ,
\label{tranf-equat}
\end{equation} 
where $N(E,z)$ is the particle number per unit of energy and volume ($GeV^{-1}cm^{-3}$) while 
$Q(E,z)$ is the particle source function representing the corresponding production rate 
(in $GeV^{-1}cm^{-3}s^{-1}$). The energy loss rate, $b(E)=dE/dt$, contains all the cooling 
mechanisms discussed before so that $b(E)=-Et_{loss}^{-1}$. 

Moreover, $t^{-1}$ corresponds to the rate at which the number of particles decreases, 
either because of escaping from the jet or because of decaying so that 
$t^{-1}=t_{esc}^{-1}+t_{dec}^{-1}$. The escape rate is $t_{esc}^{-1}=c/(z_{max}-z_0)$, 
with $z_{max}-z_0$ being the length of the acceleration zone.

The general solution of the differential equation (\ref{tranf-equat}) is 
\begin{equation}
N(E,z)=\frac{1}{\mid b(E) \mid}\int_{E}^{E_{max}} Q(E',z)e^{-\tau (E,E')}dE' \, .
\end{equation}
It is worth mentioning that, since Eq. (\ref{tranf-equat}) holds for protons, pions and 
muons, a system of three coupled equations is required to be appropriately solved in 
order to find the distributions of the particles involved in the reaction chain (protons, 
pions, muons). Afterwards, the calculations of neutrino and $\gamma$-ray emissivities 
can be found through the source functions $Q(E,z)$.   

\subsection{Source functions for relativistic protons}

As discussed previously, a realistic source function $Q(E,z)$ for the relativistic protons is a 
power-law distribution. In the jet's rest frame, due to the Fermi mechanism combined with 
the time-independent continuity equation this is written as 
\begin{equation}
Q(E',z)=Q_0\left(\frac{z_0}{z}\right)^3E'^{-2}\, ,
\qquad Q_0=\frac{8q_rL_k}{z_0r_0^2ln(E_p^{max}/E_p^{min})} \, ,
\end{equation} 
where $Q_0$ is related to $K_0$ (see the Introduction above), and $E_{p}^{min}=1.2$ $GeV$ is the minimum 
proton energy. The maximum energy is assumed to be equal to
$E_{p}^{max}\simeq 10^7$ $GeV$. The above injection function transforms to the observer's 
reference frame as described in Ref. \cite{Ody-Smpon-2017,Ody-Smpon-2018}

\subsubsection{Pion distribution}

The pion source function is obtained by the product of the above distribution and 
the total number of $p-p$ collisions as
\begin{equation}
Q_{\pi}(E,z)=cn(z)\int_{\frac{E}{E_{max}}}^{1}N_p\left(\frac{E}{x},z\right)F_{\pi}\left(x,\frac{E}{x}\right)\sigma_{pp}^{inel}\left(\frac{E}{x}\right)\frac{dx}{x}
\label{Pion_distr}
\end{equation} 
where $x = E/E_p$ and $F_\pi (x,E/x)$ denotes the pion mean number produced per $p-p$ 
collision \cite{Kelner}. As can be implied from Eq. (\ref{Pion_distr}), the proton 
distribution is entering the integrand of the r.h.s. in order to provide the pion 
source function.

\subsubsection{Muon distribution}

In a similar manner, the mean right handed and left handed muon number per pion decay is 
integrated in the total injection function considering the CP invariance and also provided 
that $N_{\pi}(E_{\pi},z)=N_{\pi ^{+}}(E_{\pi},z)+N_{\pi ^{-}}(E_{\pi},z)$ as 
\cite{Romero, Reynoso} 
\begin{align}
Q_{\mu _R^{\pm},\mu _L^{\mp}}(E_{\mu}, z)&=\int_{E_{\mu}}^{E_{max}} 
dE_{\pi}t_{\pi ,dec}^{-1}(E_{\pi})N_{\pi}(E_{\pi},z)\mathcal{N} _{\mu}^{\pm}\Theta (x-r_{\pi})
\label{Muon_distr}
\end{align}
where $\mathcal{N} _{\mu}^{\pm}$
% $$ \mathcal{N} _{\mu}^{+}=\frac{r_{\pi}(1-x)}{E_{\pi}x(1-r_{\pi})^2} \, , \qquad 
%\mathcal{N} _{\mu}^{-}=\frac{(x-r_{\pi})}{E_{\pi}x(1-r_{\pi})^2} $$ 
represent the positive (negative) right (left) handed muon spectra, respectively. 

Furthermore, in Eq. (\ref{Muon_distr}) $x = E_\mu/E_\pi$, $r_\pi =(m_\mu/m_\pi)^2$ 
and $\Theta (y)$ the Heaviside function. We mention that, the pion decay rate is 
$t_{\pi ,dec}^{-1}=(2.6\times 10^{-8}\gamma _{\pi})^{-1}$ $s^{-1}$ which implies
that, the pion distribution is important for calculating the muon distribution.

\subsection{ Neutrino emissivity and neutrino intensity }

From the above discussion we see that, neutrinos are produced directly from pion 
decay as well as from their decay-products, i.e. charged muons ($\mu^\pm$). Thus, 
the total emissivity considers both contributions as
\begin{equation}
Q_{\nu}(E,z)=Q_{\pi\rightarrow\nu}(E,z)+Q_{\mu\rightarrow\nu}(E,z)
\end{equation}  
The first term gives the neutrino injection originating from pion decay as
\begin{equation}
Q_{\pi\rightarrow\nu}(E,z)=\int_E^{E_{max}}t_{\pi,dec}^{-1}(E_\pi) N_\pi(E_\pi,z)\frac{\Theta 
(1-r_\pi-x)}{E_\pi(1-r_\pi)} dE_\pi
\end{equation} 
($x=E/E_{\pi}$) while the second term gives
\begin{align}
Q_{\mu\rightarrow\nu}(E,z)=\sum_{i=1}^4\int_E^{E_{max}}t_{\mu,dec}^{-1}(E_\mu)N_{\mu_i}(E_\mu,z)  
\left[\frac{5}{3}-3x^2+\frac{4}{3}x^3+(3x^2-\frac{1}{3}-\frac{8}{3}x^3)h_i\right]\frac{dE_\mu}{E_\mu}
\end{align}
($x=E/E_\mu$). In the latter equation, $t_{\mu ,dec}^{-1}=(2.2\times 10^{-6}\gamma_\mu)^{-1}$ $s^{-1}$, 
and $h_3=h_4=-h_1=-h_2=1$. From the four different integrals of the latter summation, the first 
and second represent the left handed muons of positive and negative charge, respectively, while 
the third and fourth stand for the corresponding right handed ones \cite{Papav-Ody-Sinat-AHEP}. 
Finally, one may evaluate the neutrino intensity by integrating the emissivity over 
the acceleration zone \cite{Reynoso,Romero}
\begin{equation}
I_{\nu}(E)=\int_VQ_{\nu}(E,z)d^3r=\pi (tan\xi)^2\int_{z_0}^{z_{max}} Q_{\nu}(E,z)z^2dz
\end{equation}
Such calculations will be presented elsewhere \cite{Papav-Ody-Sinat-AHEP}.

\section{Results and discussion}

In the present work, one of our goals is to calculate the cooling rates and energy distributions
of all particles participating in the chain reactions of $p-p$ mechanism that takes place in
the hadronic astrophysical jets of binary stars and microquasars. Then, the energy spectra of
the produced high-energy neutrinos and gamma-rays are simulated numerically through the solution 
of the corresponding transfer equations.  

\begin{figure*}[ht] 
\centering
\includegraphics[width=.325\linewidth]{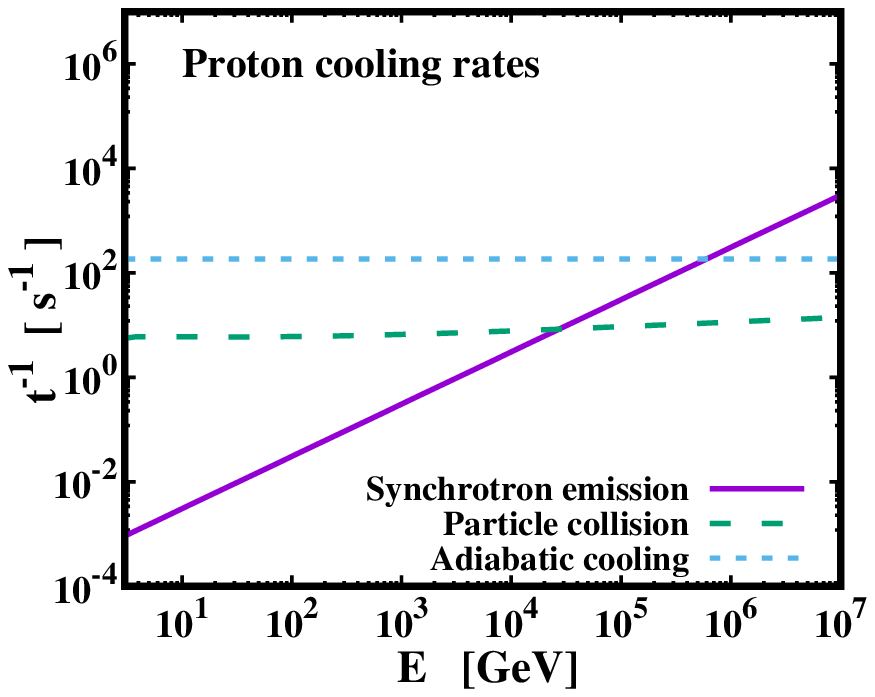}
\hspace*{-0.5 cm}
\includegraphics[width=.325\linewidth]{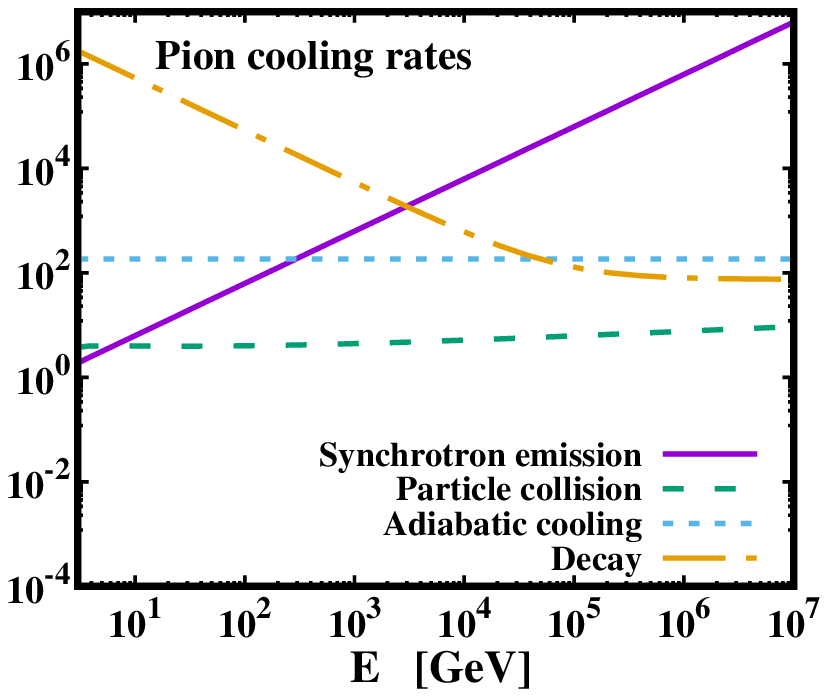}
\hspace*{-0.5 cm}
\includegraphics[width=.325\linewidth]{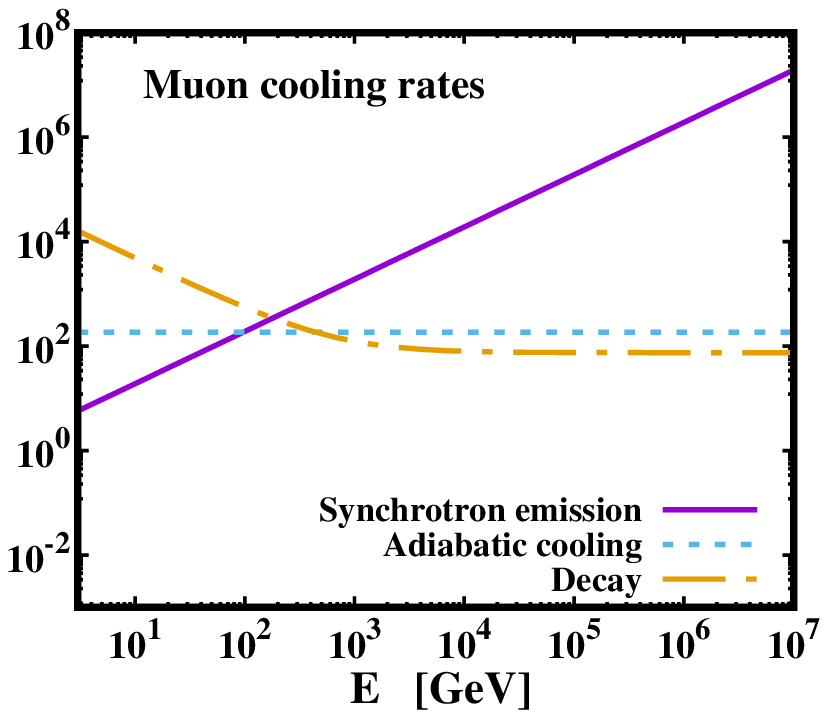} \\
\includegraphics[width=.325\linewidth]{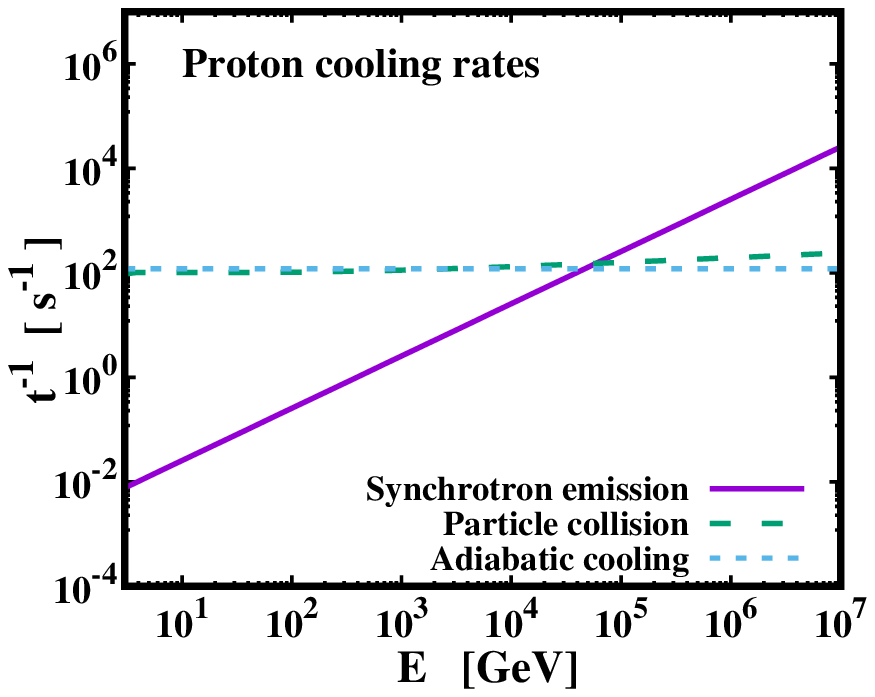}
\hspace*{-0.5 cm}
\includegraphics[width=.325\linewidth]{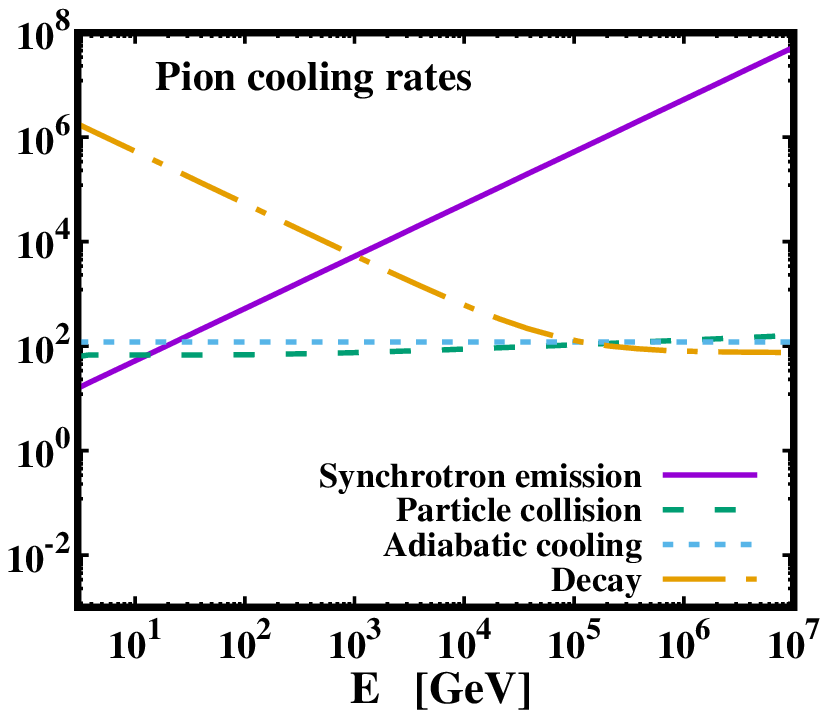}
\hspace*{-0.5 cm}
\includegraphics[width=.325\linewidth]{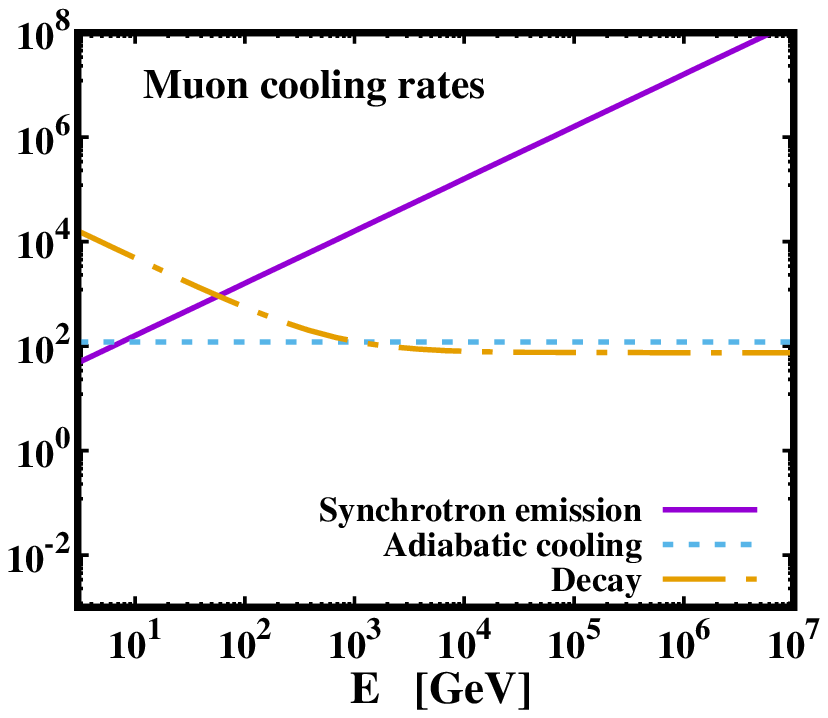}
\vspace*{0.2 cm}
\caption{\label{figure1} Cooling rates for the relativistic protons and 
the secondary particles ($\pi^\pm$, $\mu^\pm$)
produced after the p-p mechanism takes place in LMC X-1 (1st row) and Cyg X-1 (2nd row).}
\end{figure*}

By employing a C-code developed by our group here (it uses the Gauss-Legendre numerical integration 
of the GSL library), we concentrated on performing extensive calculations for the Galactic Cygnus X-1
and the extragalactic LMC X-1 binary systems. The parameter values used for LMC X-1 (and Cygnus X-1) 
are listed in Table \ref{Table1}.
By using the values of the parameters listed in Table \ref{Table1}, mostly describing geometric 
characteristics of these systems, in Fig. \ref{figure1} we display the proton, pion and muon 
cooling rates calculated for the aforementioned systems.

%%%%%%%%%%%%%%%%%%%%%%%%%%%%%%%%%%%%%%%%%%%%%%%%%%%%%%%%%%%%%%%%%%%%%%%%%%%%%%%%%%%%%%%%%%%%%%%%%
\begin{table}
\caption{\label{Table1} Model parameters describing geometric characteristics of the extragalactic 
LMC X-1, in the Large Magellanic Cloud, and the Galactic Cygnus X-1 binary systems.}
\begin{center}
\lineup
\begin{tabular}{l l l l}
\br 
Description & Parameter & LMC X-1 & Cygnus X-1 \\
\mr
Jet's base & $z_0$ & $1\times 10^8$ $[cm]$ & $1\times 10^8$ $[cm]$ \\ [0.2ex]
End of acceleration zone & $z_{max}$ & $5\times 10^8$ $[cm]$ & $5\times 10^8$ $[cm]$ \\ [0.2ex]
Mass of compact object & $M_{BH}$ & 10.91\(M_\odot\)\cite{Orosz2009} & 14.8\(M_\odot\)
\cite{Orosz2011} \\ [0.2ex]
Angle to the line-of-sight & $\theta$ & 36.38$^\circ$\cite{Orosz2009} & 27.1$^\circ$
\cite{Orosz2011} \\ [0.2ex]
Jet's half-opening angle & $\xi$ & 3$^\circ$ & 1.5$^\circ$ \\ [0.2ex]
Jet's bulk velocity & $\upsilon _{b}$ & 0.92c & 0.6c \\ [0.2ex]
\br
\end{tabular}
\end{center}
\end{table} 
%%%%%%%%%%%%%%%%%%%%%%%%%%%%%%%%%%%%%%%%%%%%%%%%%%%%%%%%%%%%%%%%%%%%%%%%%

%%%%%%%%%%%%%%%%%%%%%%%%%%%%%%%%%%%%%%%%%%%%%%%%%%%%%%%%%%%%%%%%%%%%%%%%%
%%%%%%%%%%%%%%%%%%%%%%%%%%%%%%%%%%%%%%%%%%%%%%%%%%%%%%%%%%%%%%%%%%%%%%%%

As can be seen, the particle synchrotron losses become dominant for large energies 
with the particle mass setting the separation point. In the case of protons, due to 
their large mass, synchrotron losses are not dominant up to very high energies. This
leads to a distribution with stable inclination matching the hot proton power-law 
exponent. On the other hand, for pions and muons, we see that the decay dominates 
the lower energy band and that is why the two curves in the respective graphs 
(corresponding to distributions considering energy losses along with the decay 
process only) progress similarly, especially for energies up to the decay rate 
stabilization. The synchrotron losses take off causing the smoother transition in 
the solid line's case (with energy losses) compared to the dashed one (only decay).

After calculating all the necessary distributions, the neutrino and $\gamma$-ray 
emissivities as well as the corresponding intensities may be provided. 
For the LMC X-1 system, our results have shown 
that, the increase of the half-opening angle $\xi$ leads to a decrease in the 
$\gamma$-ray production, which is an expected result since the p-p collision 
rate drops with the jet's expansion \cite{Papav-Ody-Sinat-AHEP}.

\section{Summary and Conclusions}

Black Hole X-ray binary systems (BHXRBs), consisting of a high mass compact object 
(black hole) and absorbing mass out of a companion star that results in an accretion 
disc formation, have been identified through their relativistic magnetohydrodynamical 
astrophysical flow ejection perpendicular to the aforementioned disc. This flow is 
mostly accelerated and collimated by the presence of a rather strong magnetic field 
which is initially attached to the rotational disc. 
A portion of the hadronic jet's particles (mostlye protons) 
are accelerated to relativistic velocities through shock waves travelling across 
the jet. Then, a reaction chain takes place stemming from the inelastic $p-p$ 
interactions which leads to production of neutrinos and high energy $\gamma$-rays, 
both detectable at the terrestrial extremely sensitive detectors. 

In this work, we focus mainly on the mechanisms and phenomena that affect highly the 
inelastic $p-p$ scattering and the generated secondary particles which participate 
afterwards in reaction chains leading to the emission of neutrinos and $\gamma$-rays. 
For two concrete examples, the Galactic Cygnus X-1 and the extragalactic 
LMC X-1 binary systems, we studied in more detail their cooling rates and energy 
distributions. Numerical simulations of neutrino emissivities and neutrino intensities
will be publised elsewhere. 

\section{Acknowledgments} TSK acknowledges that this research is co-financed by Greece 
and the European Union (European Social Fund-ESF) through the Operational Programme 
"Human Resources Development, Education and Lifelong Learning 2014- 2020" in the context 
of the project (MIS5047635).

\end{document}